\documentstyle[aps,prc,preprint]{revtex}
\draft
\title{Particle-hole state densities with non-equidistant
       single-particle levels}
\author{A. Harangozo, I. \c Ste\c tcu, M. Avrigeanu, and 
        V. Avrigeanu}
\address{Institute for Nuclear Physics and Engineering, P.O. Box MG-6,
         76900 Bucharest, Romania}

\begin{document}
\maketitle
\begin{center}
\vspace*{0.5in} {\sl submitted for publication in Phys. Rev. C}
\vspace*{0.25in}
\end{center}
\begin{abstract}
The correct use of energy-dependent single-particle level (s.p.l.) 
densities within particle-hole state densities based on the 
equidistant spacing model (ESM) is analysed. First, an analytical 
expression is obtained following the convolution of energy-dependent 
excited-particle and hole densities. Next, a comparison is made with 
results of the ESM formula using average s.p.l. densities for the 
excited particles and holes, respectively. The Fermi-gas model (FGM) 
s.p.l. densities calculated at the corresponding average excitation 
energies are used in both cases. The analysis concerns also the 
density of particle-hole bound states. The pairing correlations are 
taken into account while the comparison of various effects includes 
the exact correction for the Pauli exclusion principle. 
Quantum-mechanical s.p.l. densities and the {\it continuum effect} 
can also match a corresponding FGM formula, suitable for use within 
the average energy-dependent partial state density in multistep 
reaction models.
\end{abstract}

\section{introduction}

   The particle-hole state densities are basic quantities for the
description of preequilibrium emission (PE) in semiclassical models 
as well as quantum-statistical theories (e.g., 
\cite{gadioli92,feshbach92}) involving a series of particle-hole 
excitations caused by two-body interactions. The nuclear excitation in 
the equilibrium processes concerns the single-particle levels (s.p.l.) 
within an energy range of the order of the nuclear temperature around 
the Fermi level. This explains the basic role of the s.p.l. 
equidistant spacing model (ESM) \cite{ericson60} in the analysis of 
the equilibrium emission (see also \cite{bohr69}). However, much 
higher and lower single-particle energies are involved in PE reactions 
so that one should consider the reduced suitability of the ESM 
partial-state density (PSD) formula of Williams \cite{williams71}. 
Moreover, the inconsistency between the phenomenological s.p.l. 
density $g\sim A/14$ MeV$^{-1}$ and the number {\it A} of nucleons in 
the nucleus has come under increasing criticism 
\cite{herman89,chadwick91}. On the other hand, combinatorial 
calculations performed in the space of realistic shell model s.p.l. 
\cite{williams73} have other inherent shortcomings (e.g., the 
strong dependence on the basic set of s.p.l.) \cite{blann89,reffo94}. 
This accounts for the continued use of the Williams-type formula with 
various corrections \cite{fu84,kalbach8789} or exact calculation
\cite{mao93} for additional Pauli blocking and the pairing 
interaction.

In fact, there have been early attempts at considering the 
single-particle energy dependence of the s.p.l. density 
$g(\varepsilon)$ within PE formalisms 
\cite{blann7273,gadioli73,gadioli77}. Next, Kalbach 
\cite{kalbach81,kalbach85} discussed different forms of this 
dependence and found it tied to PE surface effects due to the 
interdependence of the respective assumptions. Herman {\it et al.} 
\cite{herman89} obtained an indication for the energy dependence 
nearly as that given by the Fermi-gas model (FGM) below the Fermi 
energy $F$, but linear above $F$. Chadwick and Reffo \cite{chadwick91} 
found the use of either the FGM prescription or the equidistant 
parametrization $g$=$A/F$ more accurate than the phenomenological one. 
The FGM s.p.l. density has also been involved in the development of 
the partial level densities with linear momentum \cite{chadwick91b}. 
At the same time, the ESM accuracy has been discussed in connection 
with the non-uniform s.p.l. density effect \cite{zhang92} provided by 
the harmonic oscillator model. The analysis of the energy-dependent 
s.p.l. density in the vicinity of the Fermi energy \cite{bogila92} 
provided a more general form and a good approximation of the effect 
for low energies, where the influence of the finite depth of the 
potential well can be neglected. Various $g(\varepsilon)$ have been 
obtained within both the semiclassical Thomas-Fermi approximation 
\cite{bogila92,ghosh83,blin84,hasse85b,schuck89,sato94,shlomo95,bogila96} 
and the exact quantum mechanical calculations 
\cite{dean85,shlomo92,shlomo97} which are also applicable at the high 
excitations located in the continuum region. The PSD including 
distinct energy-dependences for the excited-particle and hole level 
densities has recently been used in semiclassical \cite{avrigeanu94a} 
or quantum-statistical \cite{avrigeanu94b,avrigeanu95} cross-section 
calculations. 

The valid use of energy-dependent s.p.l. densities within the ESM 
particle-hole state density formula, even when corrected for the 
finite depth of the real nuclear potential well \cite{betak76}, has 
not yet been proved. Proving it is one aim of this work. First, the 
particle-hole state density is obtained in Sec. II by means of 
recursive relations particularly using the FGM s.p.l. density. Next, 
these are compared in Sec. III with the results of the ESM formula 
modified by using s.p.l. densities different for excited particles 
and holes, obtained from the FGM at the respective average-excitation 
energies \cite{kalbach85} (the average energy-dependent ESM 
formalism). The analysis is also carried out for the density of 
particle-hole bound states, with single-particle excitations not 
exceeding the nucleon binding energy \cite{oblozinsky86}. The advanced 
pairing correction \cite{fu84,kalbach8789} is taken into account while 
the comparison of various effects includes the exact correction for 
the Pauli exclusion principle \cite{mao93}. The importance of distinct 
corrections in the average energy-dependent ESM formalism is further 
discussed in Sec. IV. At the same time the subtraction of the free-gas 
contribution \cite{dean85,shlomo92,tubbs79} is analysed within this 
formalism, thus making no use of arbitrary truncation 
\cite{mazurek79}. The respective results are compared with the 
semiclassical and quantum-mechanical calculations of the continuum 
effect. Since the actual quantum-statistical analyses of the multistep 
reactions still involve the rough ESM, the respective results could be 
altered following consideration of the effective $NN$-interaction 
strength as the only free parameter. The conclusions are drawn in Sec. 
V.

\section{The $PSD$ recursive formula}
\subsection{Single-particle level densities}

Densities of the excited particles and holes with distinct energy 
dependences or even different values at the Fermi energy $F$ 
were considered by Gadioli and co-workers \cite{gadioli73,gadioli77},
B\v{e}t\'{a}k and Dobe\v{s} \cite{betak76}, and Herman {\it et al.} 
\cite{herman891}. The subsequent study \cite{herman89} of unperturbed 
shell-model Hamiltonian spacings indicated a linear energy dependence 
for excited particles, as well as different corresponding values at 
the Fermi level. On the other hand, Schmidt {\it et al.} 
\cite{schmidt82} found that the smooth s.p.l. density in a Woods-Saxon 
potential lies between the density corresponding to an infinite box 
and the one for an harmonic oscillator, and approximately follows 
$g(\varepsilon)\sim\varepsilon$. Moreover, this energy dependence has 
already been used within an improved abrasion model for heavy-ion 
collisions \cite{gaimard91}. 

Given the need for an analytical PLD expression, we have followed the
method of Bogila {\it et al.} \cite{bogila92} while the finite depth 
of the nuclear potential well and the case of particle-hole bound 
states have also been considered. Actually, the particle-hole bound 
state formula turns into the common form in the limit of large values 
of the nucleon binding energy $B$. The following discussion will 
concern the general form of $g(\varepsilon)$, with the $g_0$ value at 
the Fermi level. However, the usual FGM energy dependence 

\begin{equation}\label{eq:1}
  g(\varepsilon)=g_0\left({{\varepsilon}\over{F}}\right)^{1/2}
 ={{3A}\over{2F}}\left({{\varepsilon}\over{F}}\right)^{1/2} 
\end{equation}
is particularly taken into account (see also the Appendix).
This can be expressed in terms of the single-particle excitation 
energies $u=\varepsilon-F$ for particles, and $u=F-\varepsilon$ for 
holes. Next, similarly to Bogila {\it et al.} we have retained the 
first three terms of its expansion in powers of $u$ around the value 
at zero excitation energy. The general forms of the excitation-energy 
dependence then become

\begin{equation}\label{eq:2}
g_p(u) = (au^2+bu+c) \theta (B-u)
\end{equation}
and
\begin{equation}\label{eq:3}
g_h(u) = (au^2-bu+c) \theta (F-u) \: ,
\end{equation}
where the theta functions are unity if their argument is greater than
zero, and zero otherwise. The FGM values of the coefficients are
$a=g_0''/2=-g_0/8F^2$, $b=g_0'=g_0/2F$, and $c=g_0$, where $g_0'$ and
$g_0''$ are the values at the Fermi level of the respective
derivatives. Various energy dependences of the s.p.l. density can 
obviously be involved within this framework, by using the appropriate 
values for the coefficients in Eqs. (\ref{eq:2}-\ref{eq:3}).

\subsection{The convolution state-density formula}

   The bound-state density $\omega(p,h,E)$ for $p$ excited particles
above the Fermi level and $h$ holes below it ($n=p+h$), at the total 
excitation energy $E$, can be obtained by convolution of the 
single-particle and hole level densities with an excitation-energy 
conserving delta function \cite{feshbach92,chadwick91,gadioli73} 

\[
 \omega(p,h,E) = {{1}\over{p!h!}} \int_0^{\infty}du_1 g_p(u_1)
                     \int_0^{\infty}du_2 g_p(u_2) \, \ldots \,
                     \int_0^{\infty}du_p g_p(u_p)
\]
\begin{equation}\label{eq:4}
 \hspace*{1.5in} \times \int_0^{\infty}du_1 g_h(u_1) \, \ldots \, 
                 \int_0^{\infty}du_h g_h(u_h) \delta 
 \left(E-\sum_{\lambda =1}^pu_{\lambda}-\sum_{j=1}^hu_j \right) \: ,
\end{equation}
where the Pauli principle is not yet taken into account. One way to
proceed \cite{feshbach92} is to replace the $\delta$ function by its 
integral representation

\begin{equation}\label{eq:5}
\delta \left(E-\sum_{\lambda =1}^pu_{\lambda}-\sum_{j=1}^hu_j\right)=
       {{1}\over{2\pi}} \int_{-\infty}^{\infty} \exp\left[ik \left(E-
  \sum_{\lambda =1}^pu_{\lambda}-\sum_{j=1}^hu_j\right)\right]dk \: ,
\end{equation}
so that 
\begin{equation}\label{eq:6}
 \omega(p,h,E)={{1}\over{2\pi p!h!}} \int_{-\infty}^{\infty}e^{ikE}
                   \left( \int_0^{\infty} g_p(u)e^{-iku}du \right)^p
           \left( \int_0^{\infty} g_h(u)e^{-iku}du \right)^h dk \: .
\end{equation}
By using the s.p.l. densities given by Eqs. (\ref{eq:2}-\ref{eq:3}),
evaluation of the integrals, and expansion of the respective results,
it results 

\begin{equation}\label{eq:7}
 \omega(p,h,E) = {{g_0^n}\over{2\pi p!h!}}\sum_{\lambda =0}^p
  \sum_{j=0}^h(-1)^{\lambda+j}C_p^{\lambda}C_h^j R_{\lambda j}
  \left(\int_{-\infty}^{\infty}{{\exp[ik(E-\lambda B-jF)]}
  \over {(ik)^{N}}}dk\right) \: ,
\end{equation}
where, by replacing $\lambda$ by $i$, 
\begin{mathletters}\label{eq:8}
\[
R_{ij}(z)= \sum_{k_1=0}^{p-i} \sum_{i_1=0}^{k_1} 
	\sum_{l_1=0}^{h-j} \sum_{j_1=0}^{l_1}
  (-1)^{j_1+l_1}C_{p-i}^{k_1}C_{k_1}^{i_1}C_{h-j}^{l_1}C_{l_1}^{j_1}
  \left({{g_0''}\over{g_0'}}\right)^{i_1+j_1}
  \left({{g_0'}\over{g_0}}\right)^{k_1+l_1}
\]
\[
 \times \sum_{k_2=0}^{i} \sum_{i_2=0}^{k_2} 
	\sum_{l_2=0}^{j} \sum_{j_2=0}^{l_2}
 C_{i}^{k_2}C_{k_2}^{i_2}C_{j}^{l_2}C_{l_2}^{j_2}
 {{(g_0'')^{i_2+j_2}}\over{(g_0)^{i+j}}}
\]
\begin{equation}\label{eq:8a}
\times
\left(g_0''\frac{B^2}{2}+g_0'B+g_0\right)^{i-k_2}
\left(g_0''B+g_0'\right)^{k_2-i_2}
\left(g_0''\frac{F^2}{2}-g_0'F+g_0\right)^{j-l_2}
\left(g_0''F-g_0'\right)^{l_2-j_2} z \: ,
\end{equation}
$z$ is the integral which forms the functional argument in Eq. 
(\ref{eq:7}), and

\begin{equation}\label{eq:8b}
 N = n+i_1+j_1+k_1+l_1+i_2+j_2+k_2+l_2 \: .  
\end{equation}
\end{mathletters}

Finally, by using the Cauchy residue theorem we have obtained the 
expression

\[
  \omega(p,h,E)={{g_0^n}\over{p!h!(n-1)!}}
  \sum_{i=0}^p\sum_{j=0}^h(-1)^{i+j}C_p^iC_h^j
  (E-iB-jF)^{n-1}\theta(E-iB-jF)
\]
\begin{equation}\label{eq:9}
\times R_{ij}\left((E-iB-jF)^{N-n} {{(n-1)!}\over{(N-1)!}}\right) \: .
\end{equation}
The $(n-1)$ factorial and power have additionally been included in 
order to obtain a form similar to those obtained by Oblo\v zinsk\' y 
\cite{oblozinsky86} in the frame of the ESM, and Bogila {\it et al.} 
\cite{bogila92} within the FGM with no constraints for particles or 
holes. Thus, in the limiting case of large $B$ and $F$ in the above 
expression, the $i$ and $j$ indices become zero so that the last 
four sums in $R_{ij}$ disappear and the formula of Bogila {\it et al.} 
is obtained

\[
 \omega(p,h,E) = \omega^{\mbox{E}}(p,h,E) 
  \sum_{k_1=0}^p \sum_{i_1=0}^{k_1} \sum_{l_1=0}^h 
   \sum_{j_1=0}^{l_1} 
  (-1)^{j_1+l_1} C_p^{k_1} C_{k_1}^{i_1} C_h^{l_1} C_{l_1}^{j_1}
\]
\begin{equation}\label{eq:10}
 \hspace*{2.0cm} \times 
  \left({{g_0''}\over{g_0'}}\right)^{i_1+j_1}
  \left({{g_0'}\over{g_0}}\right)^{k_1+l_1}            
         E^{i_1+j_1+k_1+l_1}{{(n-1)!}\over{(n-1+i_1+j_1+k_1+l_1)!}}
         \: ,
\end{equation}
where 

\begin{equation}\label{eq:11}
 \omega^{\mbox{E}}(p,h,E) = {{g_0^nE^{n-1}} \over {p!h!(n-1)!}}
\end{equation}
is the well-known Ericson formula \cite{ericson60} for the ESM case. 
However, we underline that a definite single-particle ground state 
should be marked out by the finite value of $F$ within a consistent 
energy-dependent s.p.l. density. Therefore, the finite depth of the 
nuclear-potential well should be explicitly present in the 
particle-hole state density formulas exceeding the ESM framework. The 
usual formula for the density of the particle-hole bound states within 
the ESM approximation results immediately from Eq. (\ref{eq:9}) by 
noting the unity values of all functionals $R_{ij}$ for a constant 
s.p.l. density. The only difference with respect to the formula of 
Oblo\v{z}insk\'{y} \cite{oblozinsky86} concerns the Pauli-blocking 
factor $A_{ph}$ and the corresponding minimum energy $\alpha_{ph}$ for 
a $p$-$h$ state, which have not yet been included here. On the other 
hand, the temporary omission of the Pauli-principle correction can be 
used to estimate the energy-dependence effect better. Thus, the ratio 
between the results of Eq. (\ref{eq:9}) and Oblo\v{z}insk\'{y} formula 
without the Pauli correction is shown in Fig. 1 for a few simple 
$p$-$h$ configurations. Here the bound-state condition is released, 
$g_0$=14 MeV$^{-1}$, and $F$=38 MeV. These particle-hole states are 
actually the most important ones for the PE description, and are least 
affected by the Pauli-principle oversight. Also shown is the ratio of 
the PSDs given by Eq. (\ref{eq:9}) and the Ericson formula, in 
agreement with the trend of the Bogila {\it et al.} results (Fig. 1 of 
Ref. \cite{bogila92}) if one takes into account the different $F$ 
values used in these analyses. While the former ratio describes the 
energy-dependence effect versus an ESM formula including the potential 
finite-depth, the latter has the same role versus the simplest ESM 
expression of Ericson. 

First, the case of the $1p1h$ configuration, Fig. 1(a), shows that 
the deviation of $g_h(u)$ from $g_0$ is not really compensated by the 
corresponding deviation of $g_p(u)$, within the s.p.l. density 
convolution. Thus, the Oblo\v zinsk\' y and Ericson formulas provide 
the same PSD at excitation energies below $F$, which are higher 
than the results of Eq. (\ref{eq:9}). Second, a consideration of the 
potential finite depth decreases the Oblo\v zinsk\' y PSD values 
versus the Ericson formula. At the same time, the results of Eq. 
(\ref{eq:9}) decrease less significantly above $E$=$F$ because of the 
deviation of $g_p(u)$ from $g_0$. Moreover, the case of the $2p1h$ 
configuration shows the increased importance of the energy dependence 
versus the Oblo\v zinsk\' y formula, with the deviation from unity of 
the respective ratio becoming relevant at excitations higher than $F$. 
Smaller deviations have been obtained \cite{bogila92} as compared with
the Ericson expression.

\subsection{Inclusion of Pauli blocking and pairing correction}

Following the related forms of the recursive PSD expression 
(\ref{eq:9}) and Oblo\v{z}insk\'{y} formula \cite{oblozinsky86}, the 
correction for the Pauli blocking and pairing effects can be 
implemented within the former by inclusion of (i) a $p$-$h$ 
configuration-dependent threshold energy in the theta function, and 
(ii) the Pauli-blocking and pairing-correction term of the excitation 
energy, within the Kalbach formulation \cite{kalbach8789} 

\begin{equation}\label{eq:12}
A_K(p,h)= E_{\mbox{th}}(p,h)-{{p(p+1)+h(h+1)}\over{4g_0}}+
           {{(p-1)^2+(h-1)^2}\over{g_0F(p,h)}} \: ,
\end{equation}
where the threshold energy  

\begin{equation}\label{eq:13}
E_{\mbox{th}}(p,h)= {{g_0(\Delta_0^2-\Delta^2)}\over{4}}+
     p_m\left[\left({{p_m}\over{g_0}}\right)^2+\Delta^2\right]^{1/2}
\end{equation}
is determined by the ground and excited-state gaps $\Delta_0$ and 
$\Delta(p,h,E)$. The ground-state gap is related to the condensation 
energy $C=g_0\Delta_0^2/4$ which can be given by the constant-pairing 
correction $U_p$ \cite{fu84}, based on the odd-even mass differences 
(e.g., \cite{dilg73}). $\Delta$ is obtained by using the 
parametrization \cite{fu84,kalbach8789}

\begin{equation}\label{eq:14}
 {{\Delta}\over{\Delta_0}} = [0.996-1.76(n/n_c)^{1.60}(E/C)^{-0.68}]
                             \: \theta (E-E_{phase}) \: ,
\end{equation}
where $n_c=0.792g_0\Delta_0$ is the critical number of excitons, and
$E_{phase}$ is the energy of the pairing phase transition given by

\begin{equation}\label{eq:15}
 E_{phase} =  C \, [0.716+2.44(n/n_c)^{2.17}] 
              \: \theta (n/n_c - 0.446) \: .
\end{equation}
Actually, the latter theta function has been introduced by Kalbach 
\cite{kalbach8789} in order to explicitly take into account the lack 
of a phase transition for small $n$.

The inclusion of $p_m$=$maximum(p,h)$ and the form of the second term 
in the Kalbach correction (\ref{eq:12}) have been adopted for a 
Pauli-correction function symmetric in particles and holes,
including the effects of passive holes. Next, the third term in Eq. 
(\ref{eq:12}) has been added in order to force the PSD to have the 
values of $g_0$ and $2g_0$ for $E=E_{\mbox{th}}(p,h)$ and 
$E=E_{\mbox{th}}(p,h)+1/g_0$, respectively. The function 

\begin{equation}\label{eq:16}
F(p,h)  = 12+4g_0[E-E_{\mbox{th}}(p,h)]/p_m
\end{equation}
restricts the action of this third term to just around 
$E_{\mbox{th}}(p,h)$.

  Consequently, the PSD recursive formula (\ref{eq:9}), now including 
the Pauli and pairing corrections, becomes

\begin{eqnarray} \label{eq:17}
  \omega(p,h,E) & = & {{g_0^n}\over{p!h!(n-1)!}}
  \sum_{i=0}^p\sum_{j=0}^h(-1)^{i+j}C_p^iC_h^j
  [E-A_K(p,h)-iB-jF]^{n-1}  \nonumber \\
  & \times & \theta(E-E_{\mbox{th}}-iB-jF) \:
  R_{ij}\left([E-A_K(p,h)-iB-jF]^{N-n} {{(n-1)!}\over
  {(N-1)!}}\right) .
\end{eqnarray}

The completeness of the Eqs. (\ref{eq:9}) or (\ref{eq:17}) has the 
drawback of making them difficult to use in reaction calculations, due 
to the intricate form (\ref{eq:8a}) of the functionals $R_{ij}$. 

\section{The average energy-dependent formula}

Since approximate but simpler solutions are still of real interest, 
we will discuss below the Kalbach \cite{kalbach85} attempt to use the 
energy-dependent s.p.l. density within the ESM formula. In fact, we 
will be checking its correctness against the exact expression 
(\ref{eq:17}). It seems important to note that the Kalbach approach 
involves distinct but average s.p.l. densities for the holes and 
excited particles, respectively, at average excitation energies.

\subsection{The finite-depth and pairing corrections}

   The general form of the ESM density of particle-hole bound states  
corrected for (i) the Pauli exclusion principle \cite{williams71}, 
(ii) pairing interactions \cite{fu84,kalbach8789}, and (iii) the 
finite depth of the nuclear potential well \cite{betak76,oblozinsky86} 
can be written, similarly to Kalbach \cite{kalbach85},

\begin{equation}\label{eq:18}
 \omega(p,h,E) = {{g_0^nE^{n-1}}\over{p!h!(n-1)!}} 
                   f_K(p,h,E,F) \: ,
\end{equation}
where 

\[
 f_K(p,h,E,F) = \sum_{i=0}^p\sum_{j=0}^h (-1)^{i+j}C_p^iC_h^j
              \left[{{E-A_K(p,h)-iB-jF}\over{E}}\right]^{n-1}
\]
\begin{equation}\label{eq:19}
    \times \theta(E-E_{\mbox{th}}-iB-jF) 
\end{equation}
is a function including the finite-depth, Pauli-blocking and pairing 
corrections, as well as the bound-state condition. The original 
correction function \cite{kalbach85} was used to modify the PSD 
formula for the infinite potential well, with the Pauli correction 
terms $A(p,h)$ neglected in all terms except the leading one. However, 
by means of Eq. (\ref{eq:18}) the function $f_K(p,h,E,F)$ can now be
regarded as the ratio between the actual PSD formula and the Ericson 
expression. The index $K$ is related to the inclusion of the advanced 
pairing correction of Kalbach \cite{kalbach8789}, i.e., of the terms 
$A_K(p,h)$ and $E_{\mbox{th}}$.

The importance of the Pauli-blocking and pairing correction term 
$A_K(p,h)$ is shown in Fig. 2 for simple $p$-$h$ configurations. 
First, omission of this correction provides the unit value of the 
function $f_K$ in the case of the $xp1h$ configurations at 
excitation energies lower than $F$. Second, there is a rather similar 
case (or even identical for the state $1p1h$) if the Pauli correction 
$A_{ph}$ and the corresponding minimum energy $\alpha_{ph}$ 
\cite{oblozinsky86} are taken into account but not the pairing 
corrections. A small threshold behavior becomes apparent in this case. 
Third, the inclusion of the advanced pairing correction yields a 
strong reduction at lower excitation energies. Finally, the 
bound-state condition obviously provides a different function, mainly 
determined by the number of holes.

\subsection{Single-particle average excitation energies}

  In order to take into account the long-range deviations from ESM, 
Kalbach \cite{kalbach85} proposed the use of average values for the 
FGM s.p.l. density (\ref{eq:1}) corresponding to average excitation 
energies for either particles or holes. As a first approximation, 
these energies were estimated in the ESM frame. We follow the same 
method below but also include the case of the bound states.

The probability for the occurrence of a $p$-$h$ state with the 
excitation energy $E$ and an excited particle between $u$ and $u$+$du$ 
is given by $g_p(F+u) \cdot \omega(p-1,h,E-u,F)du/\omega(p,h,E,F)$. 
Consequently, the average excitation energy per excited particle is 
given by 

\begin{equation}
 {\overline u_p} ={{1}\over{p}}  \int_0^{\tilde B}
   {{u \cdot g_p(F+u)\, \omega(p-1,h,E-u,F)}\over{\omega(p,h,E,F)}}
   \,du  \: ,
\label{eq:20}
\end{equation}
where $\tilde B=minimum(E,B)$. Assuming a slow energy dependence of 
the correction term $A_K$, the average excitation energies for either
particles or holes become

\begin{mathletters}\label{eq:21}
\begin{equation}\label{eq:21a}
 {\overline u_p} = {{E}\over{n}} {{f_K^+(p,h,E,F)}\over
                   {f_K(p,h,E,F)}}
\end{equation}
\begin{equation}\label{eq:21b}
 {\overline u_h} ={{E-p \overline u_p}\over{h}} \: ,
\end{equation}
\end{mathletters}
where

\begin{eqnarray} \label{eq:22}
 f_K^+(p,h,E,F)&=&\sum_{i=0}^p\sum_{j=0}^h (-1)^{i+j}
   C_p^iC_h^j \left[ {{E-A_K(p,h)-iB-jF}\over{E}} 
   \right]^n \nonumber \\
 & \times & \left[1+{{n}\over{p}} {{iB}\over
   {E-A_K(p,h)-iB-jF}}\right] \theta(E-E_{\mbox{th}}-iB-jF)
\end{eqnarray}
returns to $f_K(p+1,h,E,F)$ for large $B$ \cite{kalbach85}. The 
shapes of the two functions $f_K(p,h,E,F)$ and $f_K^+(p,h,E,F)$, and 
the average excitation energies for simple $p$-$h$ configurations are 
shown, bound-state case included, in Fig. 3. The above-mentioned 
similarity between the functions $f_K^+(p,h,E,F)$ and $f_K(p+1,h,E,F)$ 
can be observed, in the general case, for the $1p1h$ and $2p2h$-states 
in Figs. 3(a) and 3(b).

It is worth noting the results of Eqs. (\ref{eq:21}) for the 
bound-plus-continuum states shown in Fig. 3(e), and the bound states 
only as displayed in Fig. 3(f). The distinct trends are due to the 
separate constraints on hole excitation up to a value of $F$ in the 
former circumstance, and a particle excitation limited by the $B$ 
value in the latter. First, ${\overline u_p}$ and ${\overline u_h}$ 
increase nearly as $E/n$ at the lowest excitations, the slightly 
larger values for holes arising from the fact that $f_K^+(p,h,E,F)$ is 
smaller than $f_K(p,h,E,F)$. Next, the constrained average excitation 
energy of either holes (in the general case) or particles (for bound 
states) becomes rather saturated at energies $E$ above the values of 
$F$ and $B$, respectively. The ESM basis of Eqs. (\ref{eq:21}) 
determines the saturation values around $F/2$ for ${\overline u_h}$ 
in the former case, and around $B/2$ for ${\overline u_p}$ in the 
latter. Moreover, in the limit of the $1p1h$ configuration a quite 
sudden change can be observed at total excitation energies around the 
values of $F$ and $B$, respectively. There is also a small change in 
the trend of ${\overline u_h}$ and ${\overline u_p}$ just below the 
maximum excitation energy $E$ for a given $p$-$h$ bound-state 
configuration, due to the final occupation of the highest allowed 
single-particle levels.

To underscore the correlation between the specific shapes of the 
correction functions $f_K$, average excitation energies 
${\overline u_p}$ and ${\overline u_h}$, and corresponding average 
values of the s.p.l. densities 

\begin{mathletters}\label{eq:23}
\begin{equation}\label{eq:23a}
 g_p(p,h) = g(F+{\overline u_p})
\end{equation}
\begin{equation}\label{eq:23b}
 g_h(p,h) = g(F-{\overline u_h}) \: ,
\end{equation}
\end{mathletters}
these quantities are shown together in Fig. 4 for the configuration 
$2p1h$. First, the effect of the Pauli and pairing correction is 
rather small. The saturations of ${\overline u_p}$ and 
${\overline u_h}$ are distinctly caused by the bound-state condition, 
Fig. 4(c), and finite-depth correction, Fig. 4(d), respectively. This 
is why the energy dependence of $g_p(p,h)$, Fig. 4(e), is distinct 
from the one of $g_h(p,h)$, Fig. 4(f), either in the general case 
(solid curves) or for the particle-hole bound states (dotted-dashed 
curves).

\subsection{ESM formula with average energy-dependent s.p.l. 
              densities}

   The average energy-dependent formula was finally obtained by using 
the average s.p.l. densities (\ref{eq:23}) within the PSD formula 
(\ref{eq:18}) which becomes

\begin{equation}\label{eq:24}
 \omega(p,h,E)={{[g_p(p,h)]^p[g_h(p,h)]^hE^{n-1}}
               \over{p!h!(n-1)!}}f_K(p,h,E,F) \: .
\end{equation}
It approximately takes into account the energy dependence of the 
s.p.l. density, even though the simple ESM form is still in use.
However, there is no basic argument why this procedure should be used 
so that its accuracy needs further study.

The method adopted in this respect consists in a comparison of the
results obtained by means of the approximate formula and the recursive 
Eq. (\ref{eq:17}). The corresponding predictions and their ratio are 
shown in Fig. 5 for the same simple $p$-$h$ configurations. The global 
values $F$=38 MeV, $B$=10 MeV, $g_0$=14 MeV$^{-1}$, and $\Delta_0$=1 
MeV were used. First of all, in the general case of the particle-hole 
state densities (i.e., for $B \rightarrow \infty$) there is only a 
small difference between the two PSD formulas even at medium energies. 
The agreement improves for more complex configurations, where the 
average of the single-particle excitation energies becomes really 
meaningful. Similar agreement is seen when the respective densities 
for the particle-hole bound states are compared within the first half 
of the energy range for each $p$-$h$ configuration. However, the 
difference becomes significant near the maximum energy for a given 
particle-hole state, and increases as the PSD values drop back to 
zero. Nevertheless, the disagreement of these bound-state density 
formulas at the high-energy extremity should have little or no effect 
on the reaction cross-section calculations.

Therefore we may conclude that the results obtained by using the 
average energy-dependent s.p.l. densities within the ESM formula are 
rather close to the exact convolution of the energy-dependent s.p.l. 
density. The next question concerns the need for this average
energy-dependent approach. The answer can be obtained by comparing the 
average energy-dependent ESM results with the PSDs given by the widely 
used ESM formula \cite{oblozinsky86}. This is shown in Fig. 6, where 
the above global parameters were used. The pairing effect is apparent 
within this latter comparison, especially at lower energies. Besides 
this aspect, the behavior shown at medium energy is similar to the 
comparison of the recursive and Oblo\v{z}insk\'{y} formulas in Fig. 1. 
The overall difference obviously exceeds the variation between the 
predictions of the PSD recursive formula and the average 
energy-dependent ESM formalism. It is rather small for the general 
case of the particle-hole states, Figs. 6(a) and 6(c), but larger for 
the bound states, Figs. 6(b) and 6(d). This strong effect following 
the consideration of the energy-dependent s.p.l. density is 
particularly caused by the constant increase in hole excitation for 
larger $E$, and the related significant decrease of the hole-state 
density as shown in Fig. 4(f).

\subsection{Effect of exact Pauli-correction calculation}

The Pauli-exclusion effect on particle-hole state densities has 
already been subject to additional investigation by Zhang and Yang 
\cite{zhang88} who used an exact method. Kalbach established later 
\cite{kalbach8789} that no conflict exists between their results and 
the frequently used Williams formulas if the energy-dependent Pauli 
term included pairing and passive-hole effects. The ESM derivation of 
PSD formulas without any approximation in the Pauli correction term 
was performed by Baguer {\it et al.} \cite{baguer89} and Mao Ming De 
and Guo Hua \cite{mao93}. The latter extended the method to the case 
of the finite-depth potential and bound states, and included the 
Kalbach \cite{kalbach8789} pairing correction. The effect of the 
alternative use of the approximate and exact Pauli-correction, and 
the one caused by the average energy-dependent ESM formula are 
compared below.

The results that were obtained by means of Eq. (\ref{eq:18}) and
according to the exact Pauli-correction formalism \cite{mao93} are 
shown together in Fig. 7, for the most sensitive low-energy region. 
The analysis was first carried out without pairing correction, i.e., 
for $\Delta_0$=0 as shown in Figs. 7(a) and 7(c). The global value 
$g_0$=14 MeV$^{-1}$ adopted by Fu and Kalbach \cite{fu84,kalbach8789} 
was used. Then, the PSD calculation with pairing correction 
corresponding to the value $\Delta_0$=1 MeV, Figs. 7(b) and 7(d), 
completed the analysis of these effects under any circumstances.

The ratio of the PSD obtained by using the two formalisms, Figs. 7(c)
and 7(d), shows more exactly that a close agreement -- even within 
1 \% -- is established just above the threshold for each $p$-$h$ 
configuration. Minor deviations exist only for larger number of 
excitons, of less interest for multistep reaction calculations which 
include them in the so-called $r$-stage \cite{feshbach92}. Therefore, 
the results of the exact Pauli-blocking effect calculations are 
closely related to those obtained by using the approximate Pauli 
correction \cite{williams71,betak76,oblozinsky86}. The inclusion of a 
suitable pairing correction seems more significant with the following 
additional remark. The PSD for the very-few-exciton configurations 
become rather saturated within a short energy range above the 
threshold. On the other hand, these configurations have the main role 
in the description of multistep reactions. Thus, the adequate account 
of the pairing effect may be found unessential for some analyses. The 
analysis of the high-energy limit of the particle-emission spectra, 
however, is quite sensitive to both pairing and nuclear-shell effects 
(e.g., \cite{avrigeanu94a,avrigeanu95,watanabe95}).

Finally, the comparative analysis of the effects illustrated in Figs.
6 and 7 shows, on a common basis, the higher importance of the 
s.p.l.-density energy dependence versus the exact calculation of the 
Pauli correction. 

\section{realistic and global results}

The importance of the various approximations involved in the 
derivation of the PSD formulas should be well known in order to avoid 
useless effort or deficient results. First, one might want to know the 
consequences of the present analysis on the total state densities
obtained as the sum of all PSD for allowed particle-hole numbers 
$p$=$h$. Next, one might question the usefulness of the PSD formulas 
discussed above, while quantum-mechanical calculations concentrating 
on the continuum region are being developed \cite{shlomo92,shlomo97}. 

\subsection{Effect on total state density}

Average s.p. excitation energies are shown in Fig. 8 for 
representative $p$-$h$ configurations of the PSDs sum defining the 
total state density $\omega(E)$. The same global values $g_0$=14 
MeV$^{-1}$, $F$=38 MeV and $\Delta_0$=1 MeV were used as above. Thus, 
it becomes apparent that all significant terms of this sum are 
characterized by rather equal ${\overline u_p}$ and ${\overline u_h}$ 
increasing nearly as $E/n$. The saturated average excitation energy of 
the holes plays a major role in the case of the few-exciton states 
which are vital for PE description, but not for configurations around 
the most probable exciton number ${\overline n}$ \cite{fu86}. The 
latter class of configurations mainly determine the total state 
density value, so that the corresponding ESM predictions are 
meaningful in this respect. Nevertheless, the average energy-dependent 
approach should be considered for the calculation of the PSD involved 
in the first stages of the multistep processes.

\subsection{Consideration of the continuum effect}

Shlomo \cite{shlomo92} performed an exact quantum mechanical 
calculation of the s.p.l. density as the sum of the bound and 
continuum contributions in the case of finite potential wells. A 
distinct point of this approach has been the consideration of the 
free-gas states counted by the s.p.l. density for a finite potential 
well. The density of these states was calculated and subtracted by 
using Green's functions associated with the respective single-particle 
Hamiltonians. Then, the commonly used semiclassical approximations for 
the s.p.l. density were similarly considered for some widely used 
mean-field potentials. Thus, Shlomo found by means of both classes of 
methods that, for a realistic finite depth well, the s.p.l. density 
decreases with energy in the continuum region {\it (the continuum 
effect)}.

This effect may have a twofold meaning for the multistep reaction
calculations. First, the continuum s.p.l. density following the 
subtraction of the free-gas contribution should be added to the 
particle-hole bound state density. The latter quantity has been used 
for the description of the multistep compound (MSC) processes 
\cite{feshbach80}. It has been assumed to be zero outside the nuclear 
well, which is now considered less appropriate \cite{shlomo95}. 
Second, in the opinion of Bogila {\it et al.} \cite{bogila96} the 
subtraction of the free gas spectrum should be involved in all PSD for 
PE calculations. This point could be most important in accounting for 
the multistep direct (MSD) processes \cite{feshbach80} which currently 
take into account all particle-hole states. Thus a correct yet simple 
method to estimate the s.p.l. density including the continuum effect 
is needed.

The continuum effect can be taken into account within the average
energy-dependent ESM formula (\ref{eq:24}) by using a form similar to 
Eq. (30) of Ref. \cite{shlomo97} (see the Appendix) for the 
excited-particle level density. According to Eq. (\ref{eq:23a}) it 
becomes 

\begin{equation}\label{eq:25}
 g_p(p,h) = {{3A}\over{2F}}
  \left[\left(1+{{\overline u_p}\over{F}}\right)^{1/2}-
  \left({{{\overline u_p}-B}\over{F}}\right)^{1/2} 
  \theta({\overline u_p}-B)\right] \: .
\end{equation}
To emphasize the origin of the particular behaviour of 
$g_p(p,h)$, this is shown at the same time as the corresponding 
${\overline u_p}$ for the basic $2p1h$ configuration in Figs. 9(b) and 
9(d). It is obvious that the average excitation energy is unchanged 
(see Fig. 4) so that the continuum effect fully determines the 
corrected s.p.l. density for excited particles. The comparison with 
the similar quantities ${\overline u_h}$ and $g_h(p,h)$ (the same as 
in Fig. 4) demonstrates the importance of this effect, the average 
s.p.l. density becoming even lower for excited particles than for 
holes.

On the other hand, it seems worth comparing the specific average 
excitation energies and s.p.l. densities which determine the PSDs  
including the continuum effect [Figs. 9(b) and 9(d)], with the related 
quantities for the particle-hole bound states [Figs. 9(a) and 9(c)]. 
The reasons for deviations from the PSD general trend, for the two 
main additional classes of particle-hole state densities with variant 
characteristics are outlined in this way. Thus, the average excitation 
energies, limited by the value $B$ for the excited particles, 
entirely determine the s.p.l. densities for the particle-hole bound 
state densities. Actually one may note that $g_p$ has a rather 
constant value in this case. The value of $g_h$ is similarly constant 
but obviously lower in the latter case, i.e. of the bound and 
continuum state density including the continuum effect. However, the 
above-discussed aspect of $g_p$ is the key quantity for the second 
class of modified PSDs.

Moreover, these two classes of PSDs with various restrictions [Figs.
10(a) and 10(b)] are at the same time compared with the predictions of 
Eqs. (\ref{eq:23}) and (\ref{eq:24}) including only the finite-depth 
correction, also shown in Fig. 10(b). A few configurations significant 
in PE calculations are used in this respect. The ratios of each of the 
two variant PSDs to the general PSD values given in Fig. 10(b) are 
further shown in Figs. 10(c) and 10(d). It follows that at medium 
energies the size of the continuum effect on the PSD values, which is 
given by the latter class of ratios, is rather similar to that for the 
bound-state condition. Therefore, a possible replacement within MSC 
calculations of the particle-hole bound state density by the PSD 
corrected for the continuum effect \cite{shlomo95} would not 
be trivial. A similar point may concern the use within the MSD 
calculations of the PSD either including the continuum effect 
\cite{bogila96} or taking into account the free-gas single particle 
levels as well.

The relation between the results of Eq. (\ref{eq:25}) and the 
quantum-mechanical (QM) calculations should be also considered before 
further use of the former in reaction calculations. Hence, following 
Shlomo {\it et al.} \cite{shlomo92,shlomo97}, the s.p.l. density was 
calculated \cite{stetcu97} by using the respective relation with 
Green's function. As an alternative to the smearing procedure, the 
imaginary part of Green's function has been calculated separately 
for the discrete and continuous states. The regular and Jost solutions 
of the radial Schr\"odinger equation are used in the continuum. The 
smooth part of the rapidly fluctuating s.p.l. density is calculated by 
means of the Strutinski smoothing procedure \cite{ross72}. The 
Woods-Saxon (WS) potential \cite{shlomo92} was considered in this 
frame as well as in the semiclassical Thomas-Fermi (TF) formula, with 
the similar results shown in Fig. 11(a). The familiar FGM shape is 
given by the TF formula with an infinite square-well (SQ) potential, 
while the corresponding finite well (FSQ) illustrates the continuum 
effect in Fig. 11(b). It should be noted that the continuum component 
of the s.p.l. density is nearly the same within either exact 
quantum-mechanical calculations with the WS potential, or TF 
approximation with either the WS or the FSQ potential wells, provided 
that the free-gas contribution is subtracted. Moreover, a similar 
trend is obtained by means of the simple FGM formula (\ref{eq:25}) 
taking into account the continuum effect. Nevertheless, the 
quantum-mechanical s.p.l. density can be related to this formula 
only for a reduced Fermi energy, e.g. ${\overline F}\simeq 20$ MeV 
\cite{avrigeanu96}. This value has been obtained as an average value 
along the trajectory of the incident projectile with respect to the 
both nuclear density and first nucleon-nucleon collision probability. 
The usual value $F$=38 MeV causes lower $g(\varepsilon_F)$ values, 
that are not consistent with the phenomenological data. 

Therefore one may use the simple FGM energy dependence, within an 
appropriate form which matches the quantum-mechanical s.p.l. 
density including the continuum effect, in the average 
energy-dependent ESM formalism. This unsophisticated yet improved
method could provide the correct PSDs for MSD/MSC calculations, in
agreement with the consideration that the highly-excited 
single-particle states are not strongly coupled to compound nuclear 
states \cite{grimes90} or partially relaxed states of composite 
nuclei formed in nuclear reactions at intermediate energies
\cite{weidenmuller64,mustafa9293}. The question is additionally made 
intricate by the recent proof of a much shorter time scale required
to reach thermal equilibration in intermediate-energy nucleon-induced 
reactions, found to be of the order of $\sim$10$^{-22}$ sec
\cite{chiba96}. Further experimental-data analyses should thus 
consider a combination of reaction models and related PSD formalisms 
as well.

\section{summary and conclusions}

  The particle-hole state density has been obtained by means of 
recursive relations, for the bound as well as bound-plus-continuum 
states. The corresponding expressions, i.e., Eqs. (\ref{eq:9}) and 
(\ref{eq:17}), can be used for various energy dependences of the 
excited-particle and hole state densities while the particular case of 
the FGM is discussed. We have underlined that consideration of the 
finite depth of the nuclear-potential well should be explicitly 
present in the particle-hole state density formulas exceeding the ESM 
framework.

Next, the results of the recursive formula are compared with the 
Kalbach \cite{kalbach85} approximation still within the ESM formula 
but using distinct average s.p.l. densities for the holes and excited 
particles respectively, at their average excitation energies. At the 
same time the Kalbach formalism is extended to the case of the bound 
states, while the pairing and Pauli-blocking effects have been 
included in all terms of the ESM correction function. The correctness 
of the average energy-dependent ESM approach is established by 
reference to the rigorous convolution (\ref{eq:9}) of the 
energy-dependent s.p.l. densities for the case of the FGM dependence. 
The difference between the predictions of the two methods is compared 
with the similar variation between average energy-dependent form and 
the standard formula \cite{betak76,oblozinsky86}, the former being 
much lower especially in the bound-state case.

  The exact calculation of the Pauli-blocking effect, which is close 
to the well-known approximate Pauli correction 
\cite{williams71,betak76,oblozinsky86}, is also discussed. Thus it is
shown on a common basis the higher importance of the s.p.l.-density
energy dependence versus the exact calculation of the Pauli 
correction. The significant role of the pairing correction is pointed 
out, while comments are made on the circumstances under which the 
adequate account of the pairing effects could indeed appear less 
than essential \cite{watanabe95}. 

The continuum effect has been considered for the case of a FGM energy 
dependence in the average energy-dependent ESM approach. The 
continuum component of the s.p.l. density is found rather similar 
using either exact quantum-mechanical calculations with the 
Woods-Saxon potential, or Thomas-Fermi approximation with WS as well 
as finite-square potential wells, provided that the free-gas 
contribution is subtracted. A similar trend is obtained by means of 
the simple FGM formula for the s.p.l. density if the continuum effect 
is taken into account. It should be noted that no arbitrary 
truncation, e.g. in the range 15-25 MeV \cite{mazurek79}, is thus 
necessary in order to take care for the continuum effect within the 
s.p.l. density account. On the other hand, since the actual 
quantum-statistical analyses of the multistep reactions use the rough 
ESM, the results following consideration of the effective 
$NN$-interaction strength as the only free parameter could be 
altered. This point is the subject of current work along with 
systematic calculations of the s.p.l. density in the continuum and 
the correlation with PE surface effects \cite{kalbach85,avrigeanu96}.

\acknowledgments
The authors are grateful to Marshall Blann, Emil Betak, Zhang 
Jingshang, and Shalom Shlomo for valuable discussions. Thanks are also
due to the referees of this work for helpful critiquing. Assistance of 
Ms. Vivien Prager with improving the manuscript is acknowledged. 
This work has been carried out under the Romanian Ministry of Research 
and Technology Contract No. 4/A13 and the Research Contract 
No. 8886/R1 of the International Atomic Energy Agency (Vienna).

\appendix
\section*{Fermi gas model s.p.l. density in continuum}

The s.p.l. density associated with a local mean field $V$ has the
following expression in the Thomas-Fermi approximation, by taking 
into account the spin degeneracy and neglecting the spin-orbit 
interactions \cite{bogila96,shlomo92,ring80}

\begin{equation}\label{eq:a1}
g^{TF}(\epsilon)=\frac{1}{2\pi^2}
	\left(\frac{2m}{\hbar^2}\right)^{3/2}
        \int \mbox{d}{\bf r}\left(\epsilon-V({\bf r})\right)^{1/2}
        \Theta(\epsilon-V({\bf r})) \: .
\end{equation} 
The single-particle energy $\epsilon$ is measured relative to the top 
of the nuclear well, in order to make a clear distinction between the 
bound states at $\epsilon\le$0 and unbound continuum states at 
$\epsilon>$0. 

For the finite well potentials, the nucleus can be imagined inside a
spherical box of radius $R$ larger than the range of $V(r)$ (see Fig. 
1 of \cite{dean85}). In the case of a square potential well of
radius $R_0$ and depth $V_0<$0, we have from Eq. (\ref{eq:a1}) 

\begin{equation}\label{eq:a2}
g_{\Omega}^{TF}(\epsilon)=\frac{1}{2\pi^2}
	\left(\frac{2m}{\hbar^2}\right)^{3/2}
	\left[\Omega_0(\epsilon-V_0)^{1/2}\Theta(\epsilon-V_0)+
              \Omega\epsilon^{1/2}\Theta(\epsilon)-
              \Omega_0\epsilon^{1/2}\Theta(\epsilon)\right] \: ,
\end{equation} 
where $\Omega_0$=4$\pi R_0^3/3$ and $\Omega$=4$\pi R^3/3$. Since the
properties of the nucleus itself do not depend on $R$ \cite{dean85},
in the case of the finite square well one has to subtract the 
contribution of free Fermi gas when $\epsilon>0$ 
\cite{bogila96,shlomo92,shlomo97} 

\begin{equation}\label{eq:a3}
   g_f(\epsilon)=\frac{\Omega}{2\pi^2} 
       \left( \frac{2m}{\hbar^2} \right)^{3/2} \epsilon^{1/2} \: .
\end{equation} 
The s.p.l. density which is thus obtained

\begin{equation}\label{eq:a4}
   g_{FSQ}^{TF}(\epsilon)=\frac{1}{2\pi^2}
	\left(\frac{2m}{\hbar^2}\right)^{3/2}\Omega_0
	\left[(\epsilon-V_0)^{1/2}\Theta(\epsilon-V_0)-
               \epsilon^{1/2}\Theta(\epsilon)\right] \: ,
\end{equation} 
has the well-known FGM form except the continuum correction term. In
terms of the single-particle energy $\varepsilon$=$\epsilon$-$V_0$
which is measured relative to the bottom of the nuclear well, it 
becomes

\begin{equation}\label{eq:a5}
   g_{FSQ}^{TF}(\varepsilon) = 
   g_0 \left[ \left({{\varepsilon}\over{F}}\right)^{1/2}-
              \left({{\varepsilon+V_0}\over{F}}\right)^{1/2} 
		\Theta(\varepsilon+V_0)\right] \: ,
\end{equation} 
where $g_0$=$g(\epsilon_F)$=$g(F)$ with reference to both notations 
used for the s.p.l. energy. Actually, the derivation of Eq. 
(\ref{eq:a4}) shows that the radius $R$ is indeed taken into account 
but $g_{FSQ}^{TF}(\epsilon)$ does not depend on it. Therefore, the 
final expression is apparently only a difference of terms calculated 
for the SQ potential well and the free-particle case, respectively, 
within an infinite spherical box with the radius $R_0$.

\newpage
\begin{center}
{\bf FIGURE CAPTIONS}
\end{center}
\begin{itemize} \itemsep 0pt \topsep 0pt \parskip 0pt
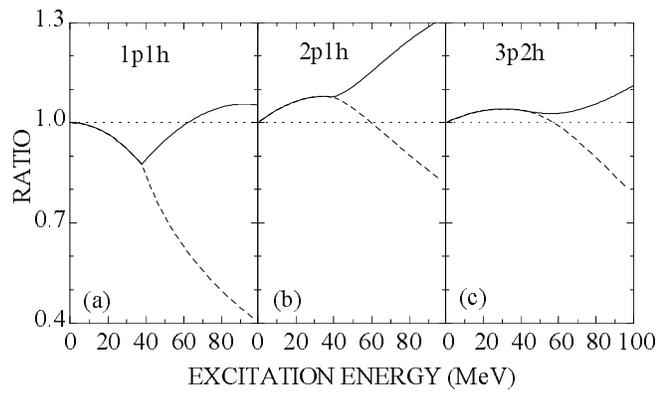
\item [FIG. 1.] The convolution state density given by Eq. 
(\ref{eq:9}), divided by the Oblo\v{z}insk\'{y} formula without Pauli 
correction (solid curves), as well as by the Ericson formula (dashed 
curves), for the given $p$-$h$ configurations. In Eq. (\ref{eq:9}) and 
the Oblo\v{z}insk\'{y} formula, $F$=38 MeV and the limit of large $B$ 
is considered. For all calculations $g_0$=14 MeV$^{-1}$.

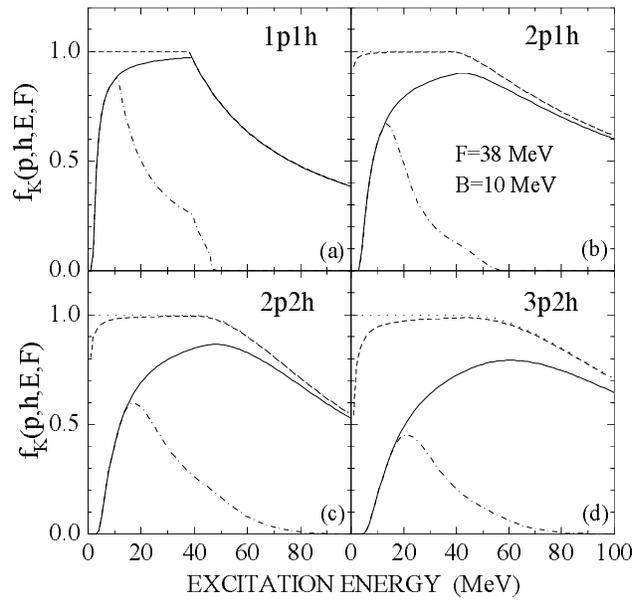
\item [FIG. 2.] The correction function for the Pauli blocking, 
nuclear-potential finite depth and pairing-correlation effects, of 
the ESM particle-hole state density for given $p$-$h$ configurations. 
The calculations use the finite-depth correction alone with $F$=38 MeV 
(dotted curves), the Pauli correction without (dashed curves) and with 
the pairing effects included (solid curves), and also the bound-state 
condition with $B$=10 MeV (dash-dotted curves).

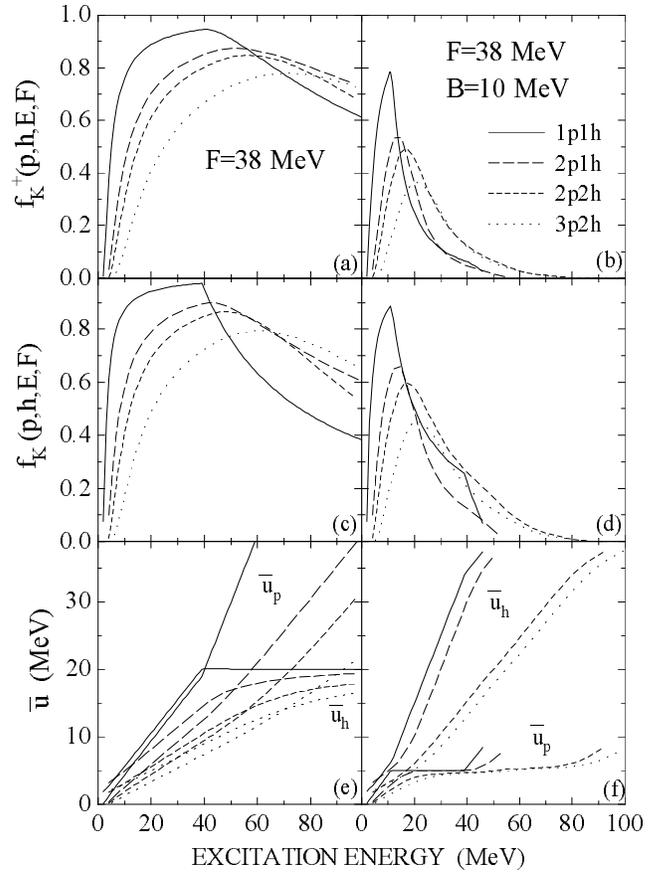
\item [FIG. 3.] (a),(b) The $f_K^+(p,h,E,F)$ and (c),(d) the 
$f_K(p,h,E,F)$ correction functions to the ESM formula for 
the Pauli blocking, potential finite-depth and pairing-correlation 
effects, as well as (e),(f) the average excitation energies for 
particles and holes within the given $p$-$h$ configurations, for 
(a),(c),(e) the particle-hole bound plus continuum states, and 
(b),(d),(f) the bound states only. For all calculations $F$=38 
MeV, and $B$=10 MeV is considered for the bound states.

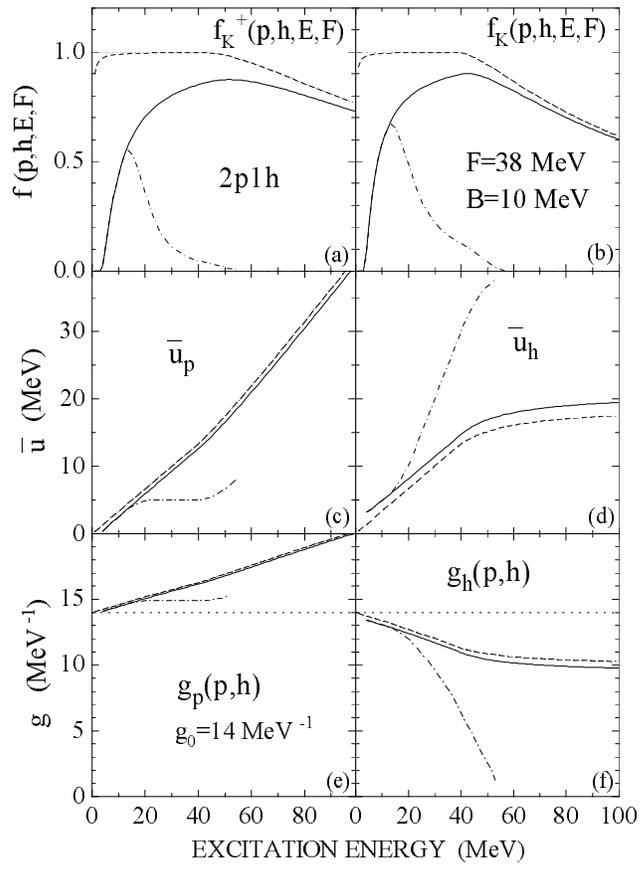
\item [FIG. 4.] (a) The $f_K^+(p,h,E,F)$ and (b) 
$f_K(p,h,E,F)$ correction functions to the ESM formula for 
the Pauli blocking, potential finite-depth and pairing-correlation 
effects, (c),(d) the average excitation energies for particles and 
holes, and (e),(f) the average energy-dependent s.p.l. densities for 
excited particles and holes, for the $2p1h$ configuration. The meaning 
of the curves is the same as in Fig. 2 except that the dotted line in
(e) and (f) gives the value of $g_0$. For all calculations $g_0$=14 
MeV$^{-1}$ and $F$=38 MeV, while $B$=10 MeV is considered for the 
bound states.

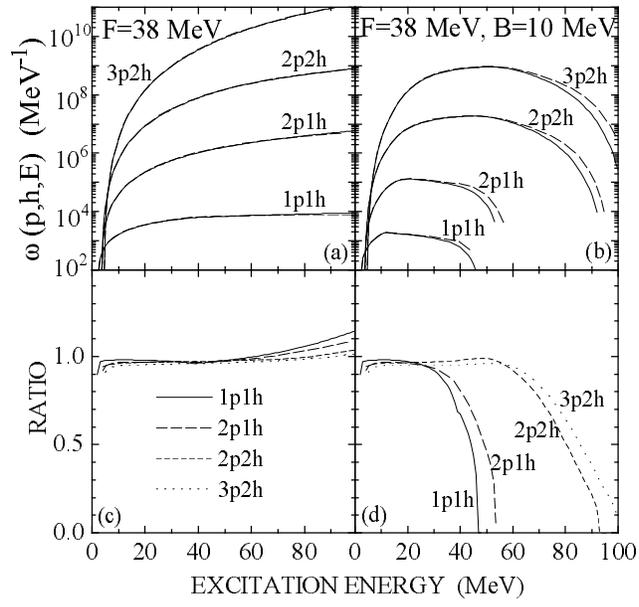
\item [FIG. 5.] The particle-hole state densities for the given 
$p$-$h$ configurations, obtained with the average energy-dependent ESM
formalism (solid curves) and the PSD recursive formula (dashed 
curves), and their ratio, for (a),(c) bound-plus-continuum states, 
and (b),(d) bound states only. The same global values are used as
in Fig. 4.

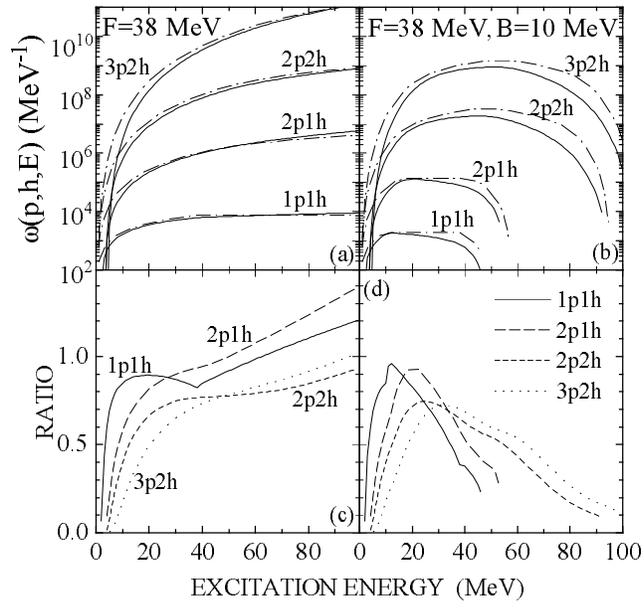
\item [FIG. 6.] The same as Fig. 5 except the latter PSD formula 
considered in the comparison is the ESM formula \cite{oblozinsky86} 
(dash-dotted curves).

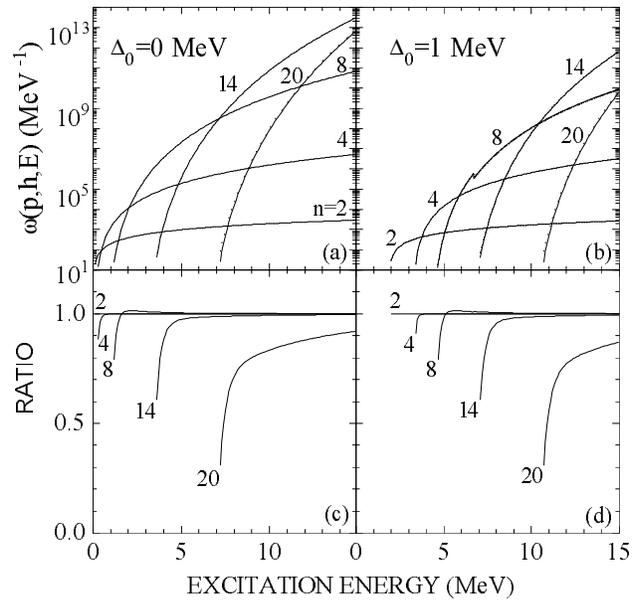
\item [FIG. 7.] (a),(b) The particle-hole state densities for given
$n$-exciton configurations with $p$=$h$, obtained with the ESM 
formulas including the advanced pairing correction \cite{kalbach8789} 
and either the exact correction for the Pauli exclusion principle 
\cite{mao93} (solid curves) or the respective approximate form 
\cite{oblozinsky86} (dashed curves), and (c),(d) their ratios. For all 
calculations $g_0$=14 MeV$^{-1}$, while $\Delta_0$=1 MeV is considered 
for the pairing correction account in (c) and (d).

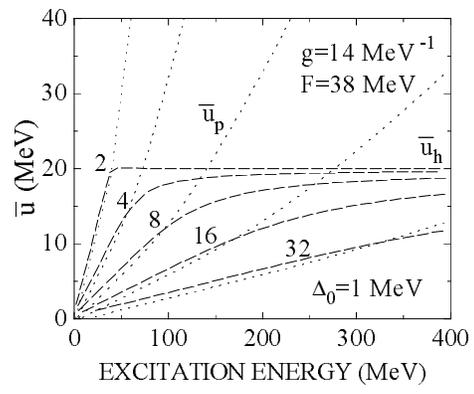
\item [FIG. 8.] Average excitation energies of the excited particles 
(dotted curves) and holes (dashed curves) for given $n$-exciton 
configurations with $p$=$h$, as function of the total excitation 
energy. The same global values are used as in Fig. 4, while 
$\Delta_0$=1 MeV is considered for the pairing correction account.

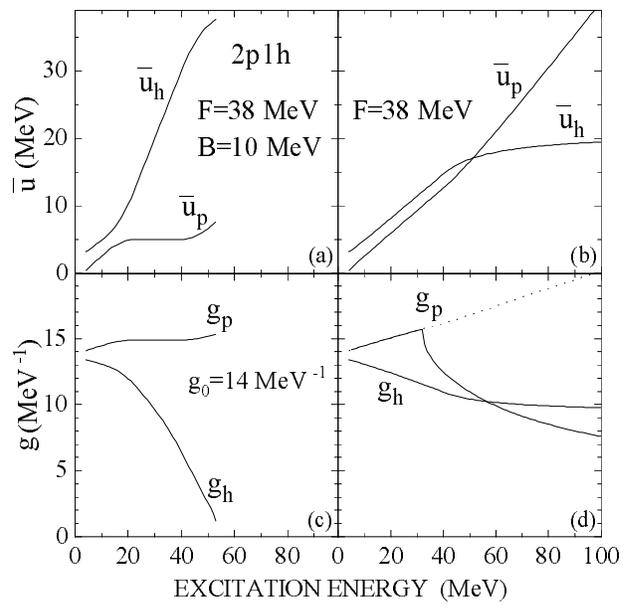
\item [FIG. 9.] (a),(b) The average excitation energies and (c),(d) 
the related s.p.l. densities for (a),(c) the particle-hole bound 
states, and (b),(d) the general case including (solid curves) 
or not (dotted curve) the continuum effect, for the $2p1h$ 
configuration. The same global values are used as in Fig. 4.

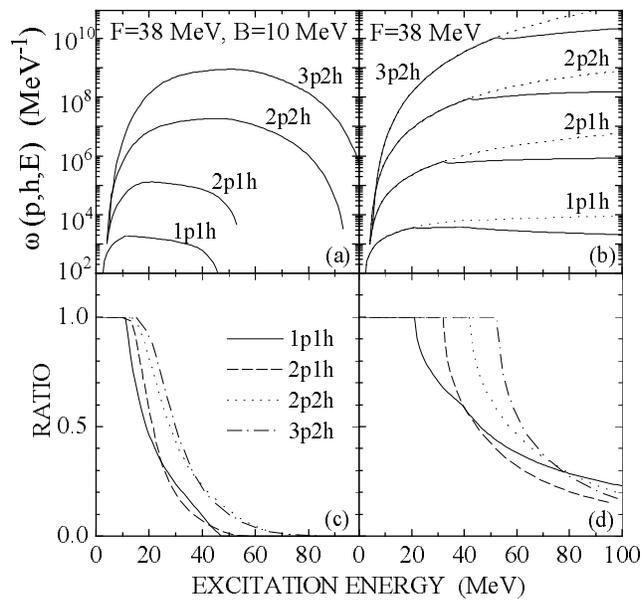
\item [FIG. 10.] The particle-hole state densities for the given 
$p$-$h$ configurations, obtained within the average energy-dependent 
ESM formalism for (a) the bound states and (b) the 
bound-plus-continuum states including the continuum effect (solid 
curves) or taking into account also the free-gas single particle 
levels as given by Eqs. (\ref{eq:23}) and (\ref{eq:24}) in the limit 
of large $B$ and with only the finite-depth correction (dotted 
curves), and (c),(d) the ratios of each of the two kinds of variant 
PSDs to the third one. The same global values are used as in Fig. 4.

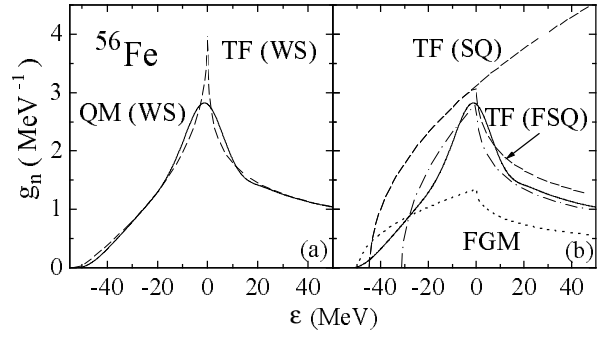
\item [FIG. 11.] The comparison of the smoothed quantum-mechanical 
s.p.l. density for the neutrons of the nucleus $^{56}$Fe (solid curve)
and the results of the TF approximation using (a) the same 
Woods-Saxon potential well as within the QM calculation, and (b) the 
infinite (SQ) and finite square potential wells (FSQ) (dashed curves), 
and the FGM formula with the Fermi-energy values of $F$=38 MeV (dotted 
curve) and $\bar F$=20 MeV (dot-dashed curve). For parameters of the 
potential wells see Refs. \cite{shlomo92,stetcu97}.

\end{itemize}
\end{document}